\documentclass[twocolumn,english]{article}
\usepackage[T1]{fontenc}
\usepackage{amssymb}
\usepackage{graphicx}

\makeatletter
\usepackage{a4}
\input{epsfig.sty}
\def\fnum@table{\tablename~{\bf\thetable}}
\def\fnum@figure{\figurename~{\bf\thefigure}}
\def\tablename{\footnotesize{\bf Table}}
\def\figurename{\footnotesize{\bf Figure}}

\def\be{\begin{equation}}
\def\ee{\end{equation}}

\makeatother

\usepackage{babel}
\begin{document}

\title{\textbf{Constraining pion interactions  at very high energies 
 by cosmic ray data}}

\author{Sergey Ostapchenko$^{1,2}$ and Marcus Bleicher$^{1,3}$\\
$^1$\textit{\small Frankfurt Institute for Advanced Studies, 
 60438 Frankfurt am Main, Germany}\\
$^2$\textit{\small D.V. Skobeltsyn Institute of Nuclear Physics,
Moscow State University, 119992 Moscow, Russia }\\
$^3$\textit{\small Institute for Theoretical Physics, Goethe-Universitat,
 60438 Frankfurt am Main, Germany}
}

\maketitle
\begin{center}
\textbf{Abstract}
\par\end{center}

We demonstrate that a substantial part of the present uncertainties in
model predictions for the average maximum depth of cosmic ray-induced
 extensive air showers 
is related to  very high energy pion-air collisions.
Our analysis shows that  the position of the 
  maximum  of the muon production profile in air showers is strongly
 sensitive to the properties of such  interactions. Therefore, the
  measurements of the maximal muon production depth
 by cosmic ray experiments provide a unique opportunity
 to constrain the treatment of pion-air interactions at very high energies
 and to reduce thereby model-related uncertainties for the shower maximum
 depth.

\section{Introduction\label{intro.sec}}
\label{sec:intro}
Experimental studies of high energy cosmic rays (CRs) are traditionally performed
using   extensive air shower (EAS) techniques: the properties 
of the primary CR particles are reconstructed from measured characteristics
of nuclear-electromagnetic cascades induced by their interactions in the
atmosphere. This naturally implies the importance of detailed Monte Carlo 
simulations of the EAS development, particularly, of its backbone -- the cascade
of nuclear interactions of both, the primary particles and
of the secondary hadrons
produced. Thus, the very success of these experimental studies depends crucially
on the accuracy of the  modeling of hadron-air collisions at high energies.
This is especially so for measurements of the  nuclear composition of 
ultra-high energy cosmic rays (UHECRs). The primary CR composition 
 is the key observable for discriminating between different
astrophysical models for the origin of the UHECRs
and is of utmost importance 
for revealing the nature of UHECR sources (for recent reviews, see
 \cite{kot11,bla14}).

Typically, one chooses between two main experimental 
procedures \cite{nagano,ung12}.
In the first case, one deals with the information obtained by scintillation
detectors positioned at ground. The energy of the primary particle is  
reconstructed from the measured lateral density of charged particles (mostly,
electrons and positrons) while the particle type is inferred from the relative
fraction of muons, compared to all charged particles at ground. Alternatively,
one may study the longitudinal EAS development by measuring fluorescence
light produced by excited air molecules at different heights in the atmosphere.
Here dedicated fluorescence telescopes are employed. In the latter
case, the primary energy is related to the total amount of fluorescence light
emitted. In turn, the particle type may be determined from the measured
position of the   shower maximum $X_{\max}$ -- the depth in the
atmosphere (in g/cm$^2$), where the number of ionizing particles reaches its
maximal value.

Not surprisingly, the observables used to determine the primary particle type -- the lateral muon density   and the EAS maximum position 
$X_{\max}$ -- appear to be very sensitive to details of high energy
hadronic interactions \cite{ulr11}. More precisely, $X_{\max}$ depends strongly on the
properties of the primary particle interaction  with air nuclei:
the inelastic cross section and the forward spectra of secondary hadrons
produced. In turn, the EAS muon content is formed in a multistep cascade
process, driven mostly by interactions of secondary pions and,
to a smaller extent, kaons with air.
 Here, we are going to demonstrate that present model predictions
for the average shower maximum depth also depend noticeably on the model 
treatment of pion-air collisions. Moreover, due to a reduction of uncertainties
related to the description of very high energy proton-proton and proton-nucleus
interactions, caused by a more reliable model calibration with the data of the
Large Hadron Collider (LHC),
the treatment of pion-nucleus collisions becomes the dominant source of 
model uncertainty concerning $X_{\max}$ predictions.  We will  demonstrate
how this can be constrained by   measurements of the   maximal   muon production depth in air showers.

\section{Uncertainties of model predictions for  $X_{\max}$}
\label{sec:model-xmax}
By far, the most suitable EAS parameter for studying primary CR composition
is the shower maximum depth $X_{\max}$. Apart from the possibility to measure
it reliably by modern air fluorescence detectors, the uncertainties of the 
respective model predictions have been greatly reduced with the start of LHC. 
Especially,  the precise measurements of
 the total and elastic proton-proton cross sections by the TOTEM and ATLAS
experiments \cite{totem13,aad14} provided strong constraints for the models.
Another potential source of uncertainty for  $X_{\max}$ is related to its
sensitivity to the rate of inelastic diffraction in proton-proton and
proton-nucleus collisions. Diffraction largely dominates the   shape of the
 very forward spectra for secondary particle production, which in turn
 makes a strong impact on the longitudinal EAS development. This has been 
investigated in Ref.\
 \cite{ost14} in the framework of the QGSJET-II-04 model \cite{ost11}, in view
 of recent   studies of diffraction at LHC.
The obtained characteristic uncertainty for  $X_{\max}$  amounted to 
 10 g/cm$^2$, being thus comparable to the  accuracy of 
the shower maximum measurements.

\begin{figure*}[t]
\centering
\includegraphics[height=5.7cm,width=0.48\textwidth,clip]{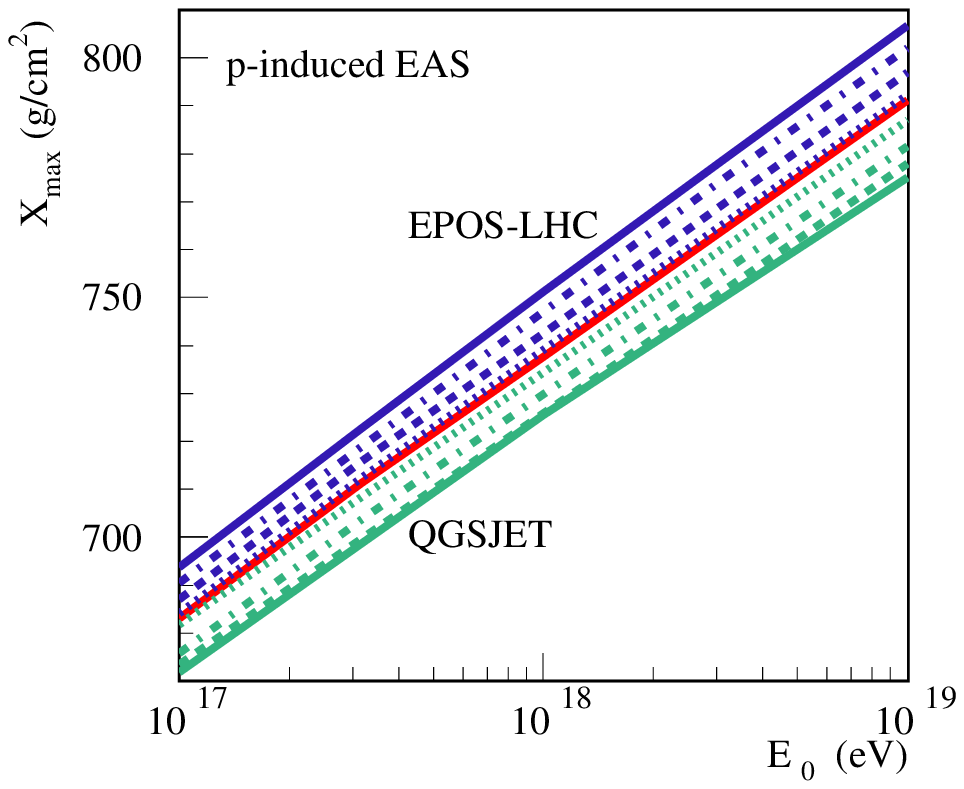}\hfill
\includegraphics[height=5.7cm,width=0.48\textwidth,clip]{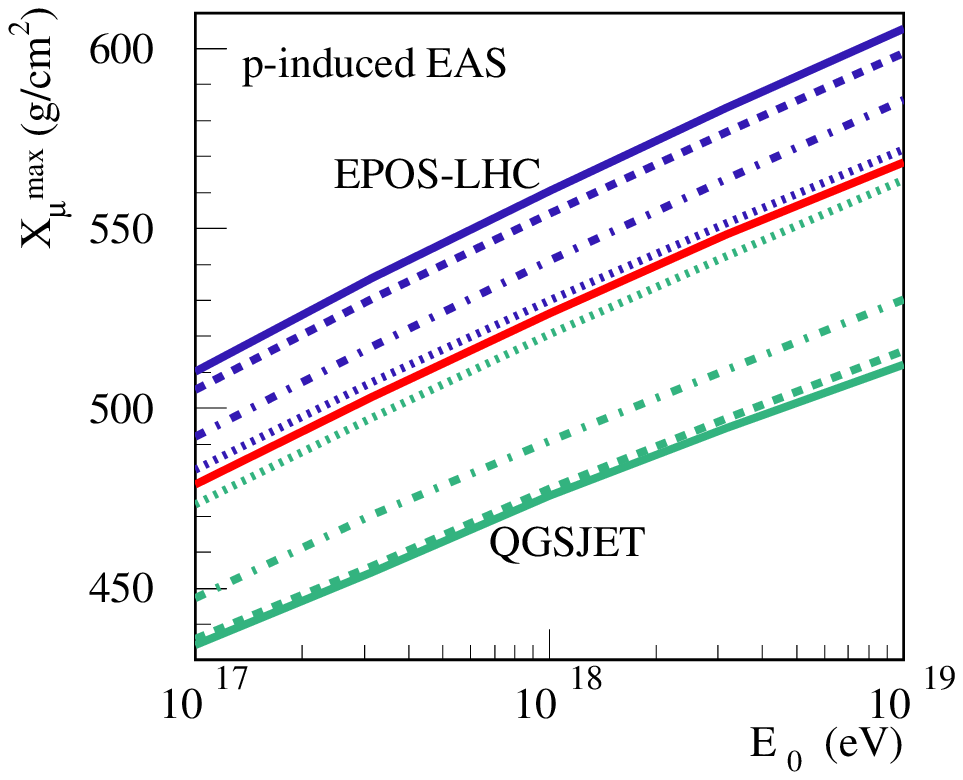}
\caption{Primary energy dependence of $X_{\max}$ (left) and of 
$X^{\mu}_{\max}$ for $E_{\mu}\geq 1$ GeV (right)
 for proton-initiated vertical EAS,
calculated using the EPOS-LHC, QGSJET-II-04, and QGSJET models (respectively
top blue, middle red, and bottom green solid lines), or applying  mixed
 model descriptions, 
as explained in the text (dashed, dotted-dashed, and dotted lines).}
\label{fig:xmax}       
\end{figure*}

However, present differences between various calculations of  $X_{\max}$ are
substantially larger, as illustrated in  Fig.\ \ref{fig:xmax} (left) %
with the 
corresponding results of   the QGSJET-II-04, EPOS-LHC \cite{pie15}, and
QGSJET \cite{qgs93,qgs97}  models.\footnote{Here and in the following 
the calculations of EAS development are performed using 
  the CONEX code \cite{ber07}.} 
Particularly
surprising is the difference between the  QGSJET-II-04 and  EPOS-LHC
predictions as both models have been recently updated using
LHC data. Thus, the question arises if the  analysis of Ref.\  \cite{ost14}
was not general enough or the position of the
  shower maximum depends on some other
characteristics of hadronic interactions, not well constrained by   present
LHC data.

To reveal the interaction features which are responsible for the above-discussed
differences in  $X_{\max}$ predictions, we are going to employ
 a ``cocktail'' model
approach: using QGSJET-II-04 to describe some selected interactions of hadrons
in the atmospheric cascades or   some particular features of the primary
interaction, while treating the rest with one of the other two 
models.\footnote{We   restrict our analysis to the case of proton-initiated
air showers: For average characteristics of nucleus-induced EAS, the 
``superposition'' model works quite well \cite{qgs93,kal89,eng92}. 
E.g., for the energy dependence of  $X_{\max}$ for iron- and proton-induced EAS, the relation  
$X_{\max}^{\rm Fe}(E_0)=X_{\max}^{p}(E_0/56)$ holds to a good accuracy.}
As the first step, we apply  QGSJET-II-04 to determine the position of the 
primary particle interaction in the atmosphere and to describe the production
of secondary nucleons in this interaction; 
all other characteristics of the first
proton-air  collision and all the subsequent interactions of secondary 
hadrons in the cascade are treated using  EPOS-LHC. This way we check the 
sensitivity of the calculated   $X_{\max}$  to the model differences
concerning the proton-air cross section and the predicted nucleon spectra,
  which thus comprise the effects of the inelastic diffraction.
The obtained  $X_{\max}$ shown by the blue dotted-dashed line in 
 Fig.\ \ref{fig:xmax} (left)   differs from the original  EPOS-LHC  results by not more
 than 5 g/cm$^2$, which is well within the uncertainty range obtained in 
 Ref.\  \cite{ost14}.
  
  Next, we apply QGSJET-II-04 to describe all the characteristics of the
  primary interaction, while treating the rest of the hadron cascade using 
  EPOS-LHC. The obtained  $X_{\max}$  shown by the blue dashed line in  
   Fig.\ \ref{fig:xmax} (left)   is shifted further towards the  QGSJET-II-04 results
   by up to  5 g/cm$^2$. This additional shift is explained by somewhat harder
   spectra of secondary mesons, most importantly, of secondary pions in  
 EPOS-LHC, compared to   QGSJET-II-04. Here we actually observe
 an important change in the physics of the hadronic cascade in the atmosphere.
 At lower energies, there is a very pronounced ``leading nucleon'' effect,
 i.e.\   most energetic secondary particles in proton-air collisions are
 typically protons or neutrons (produced either directly or via 
 decays of hyperons and resonances). On the other hand, in the very high energy
 limit the energy loss of leading nucleons is noticeably higher   
 and the most energetic secondary hadron may well be a pion or a kaon,
 which results in a stronger sensitivity of  $X_{\max}$ calculations to the
 corresponding production spectra. We also repeat the same calculation
 describing secondary hadron interactions in the cascade with   QGSJET,
 the results being plotted by the green dashed line  in   Fig.\ \ref{fig:xmax}
 (left). 
 In this case, the difference with the pure QGSJET-based calculation does not
 exceed 3  g/cm$^2$, which is due to the fact that forward particle spectra
 in proton-air collisions
 are rather similar in QGSJET and QGSJET-II-04.
 
 Thus, there remain  large differences between the two dashed lines in 
  Fig.\ \ref{fig:xmax} (left) and the results of  QGSJET-II-04, which arise
   from the model treatments of  pion- and kaon-air interactions.
 In the particular case of   QGSJET, this   amounts
 to $10-13$ g/cm$^2$, i.e.\ to $\simeq 80$\% of the difference between
 QGSJET and QGSJET-II-04, and is
 mainly related to the larger pion-air cross section  and     softer
production spectra for secondary mesons, predicted by QGSJET. 
The larger cross section   is responsible for $\simeq 20$\%
of the difference, as
is illustrated by the green dotted-dashed line in  Fig.\ \ref{fig:xmax} (left),
 obtained by using QGSJET-II-04   both for the primary
  interaction and for the inelastic cross sections
 for all the secondary hadron-air collisions
in the cascade. In turn, applying  QGSJET-II-04 to  describe also
pion and kaon spectra in pion-air collisions produces an additional $35-50$\%
effect, as shown by the green dotted line in  Fig.\ \ref{fig:xmax} (left).

In case of EPOS-LHC, the  remaining $\simeq 35$\%   difference with
 the QGSJET-II-04 results  is   both due to 
 a copious production of baryon-antibaryon pairs 
in pion-  and kaon-air collisions and due to  harder
(anti-)baryon spectra in  EPOS-LHC \cite{pie08}.   These features lead to
a slower energy dissipation from the hadronic cascade, hence, to an
elongation of the   shower profile.
Indeed, if we apply  QGSJET-II-04 to describe
both the primary interaction and the production of nucleons and antinucleons
in all the secondary pion-  and kaon-air collisions,
 while treating the rest with
EPOS-LHC,  the obtained  $X_{\max}$  
shown by the blue dotted line  in  Fig.\ \ref{fig:xmax} (left) practically
coincides with the  QGSJET-II-04 results.

\begin{figure*}[t]
\centering
\includegraphics[height=5.7cm,width=0.48\textwidth,clip]{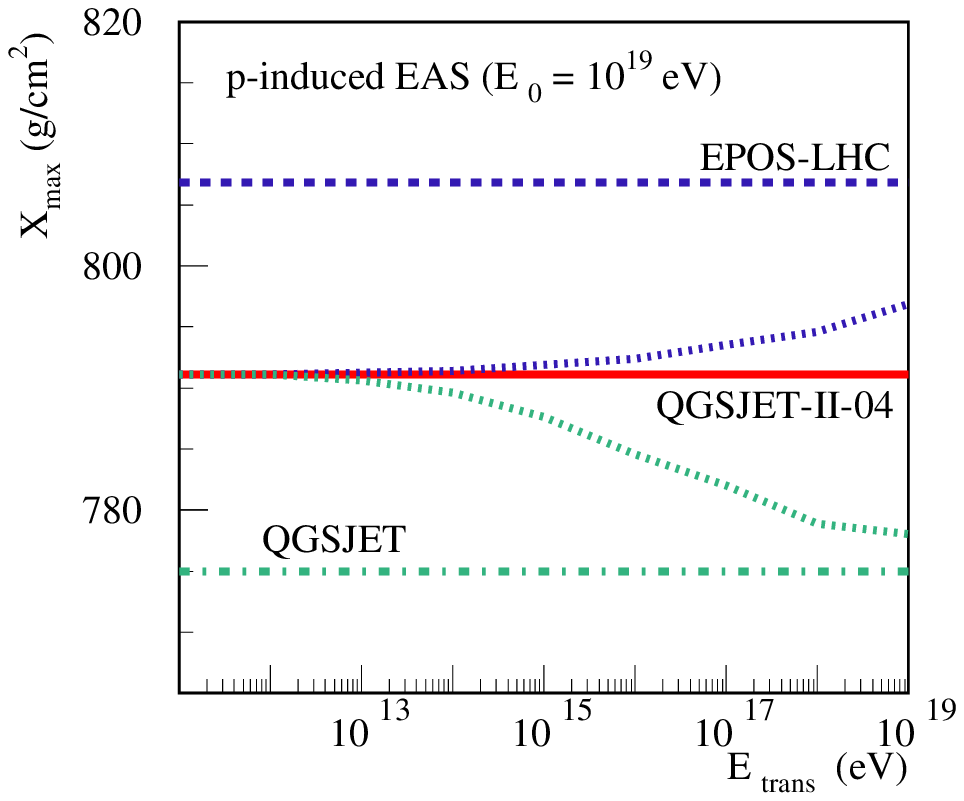}\hfill
\includegraphics[height=5.7cm,width=0.48\textwidth,clip]{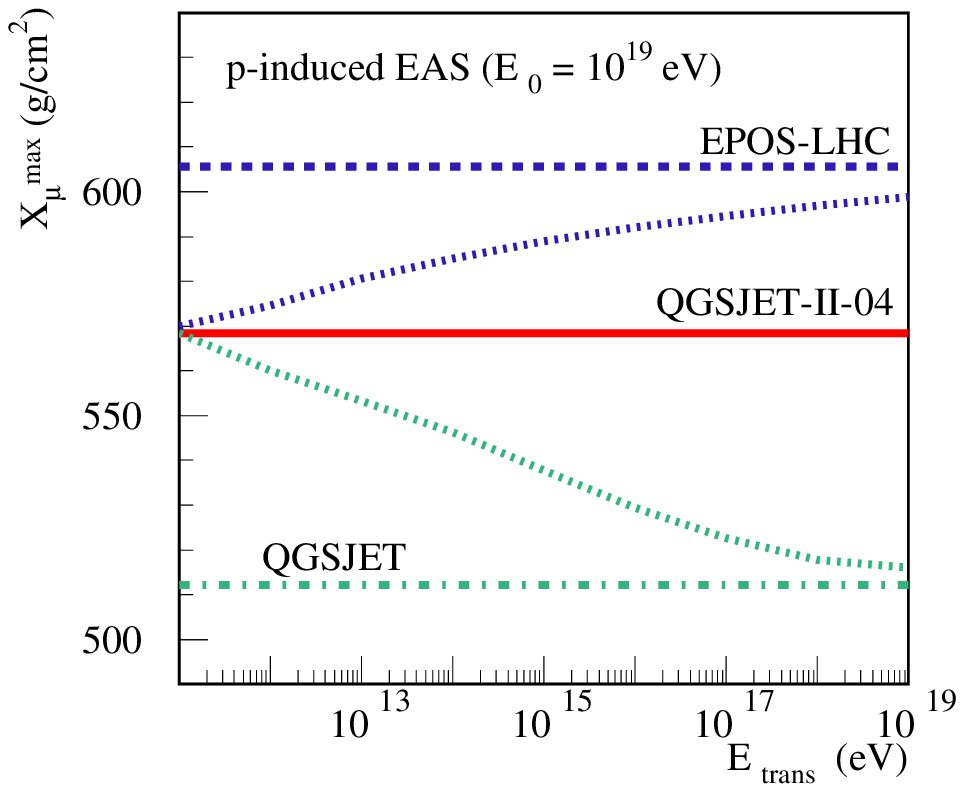}
\caption{$E_{\rm trans}$-dependence of $X_{\max}$ (left) and of 
$X^{\mu}_{\max}$ for $E_{\mu}\geq 1$ GeV (right)
 for proton-initiated vertical EAS  of energy $10^{19}$ eV,
calculated using  QGSJET-II-04 for hadronic interactions at $E>E_{\rm trans}$ 
and applying EPOS-LHC or QGSJET at  $E<E_{\rm trans}$  -
respectively blue and green dotted lines. The predictions of
the  QGSJET-II-04, EPOS-LHC, and QGSJET models
are shown by the red solid, blue dashed, and green dotted-dashed lines
 respectively.}
\label{fig:xmax-mix}       
\end{figure*}

Let us now determine the energy range of pion-  and kaon-air collisions
which are most relevant for the above-discussed model dependence of  $X_{\max}$ 
calculations. To this end, we apply QGSJET-II-04 to treat all  hadronic 
interactions in the cascade above some ``transition'' energy $E_{\rm trans}$,
while describing hadron-air collisions at $E<E_{\rm trans}$ using either
EPOS-LHC or QGSJET. The obtained dependence of the calculated  $X_{\max}$ 
on   $E_{\rm trans}$ for the two cases is shown in 
 Fig.\ \ref{fig:xmax-mix} (left) %
by respectively blue and green dotted lines  for
 $E_0=10^{19}$ eV. Not surprisingly,
the model differences for the predicted   $X_{\max}$ are due to 
very high energy pion-  and kaon-air interactions, which is reflected
in the strong $E_{\rm trans}$-dependence in the corresponding energy range. 
Indeed, as the longitudinal charged particle profile   in EAS is
dominated by the contribution of electrons and positrons,  $X_{\max}$ may 
  be influenced by pion-  and kaon-air interactions only in the
beginning of the hadronic cascade, before most of the energy of the primary
particle is channelled   into secondary electromagnetic cascades.

\section{Relation to maximal  muon production depth}
\label{sec:model-xmu}
As demonstrated in  Section \ref{sec:model-xmax}, a large part of the model
uncertainty for  the predicted  $X_{\max}$ 
is related to the
 treatment of pion-air  collisions  at very high energies.
 As no accelerator experiments with a very high energy pion beam 
  are foreseen,
this may constitute a serious obstacle for improving the accuracy of   $X_{\max}$
calculations and thus may   hamper  further progress
in experimental studies of  UHECR composition.

However, the treatment of pion-nucleus interactions can be constrained 
indirectly by studying other characteristics of  very high energy EAS.
 Recently, the Pierre Auger experiment measured 
  the   maximal muon production depth in EAS, $X^{\mu}_{\max}$ -- the depth in the atmosphere
(in g/cm$^2$), where the rate of muon production via decays of pions and kaons
reaches its maximal value \cite{pao-xmu}.
 In particular, one observed a strong contradiction between the
results of EAS simulations with EPOS-LHC  and the experimental data:
the   measured  $X^{\mu}_{\max}$ was
substantially  smaller than predicted by  that model, 
even if the  heaviest   primary CRs were considered.

 There are both similarities and differences
concerning the relation of  $X_{\max}$ and $X^{\mu}_{\max}$ to the properties
of hadron-air collisions. Obviously, both characteristics are sensitive
to the position $X_0$ of the primary particle interaction in the atmosphere,
 which depends on the respective inelastic cross section: fluctuations of  $X_0$
 shift the whole cascade upwards and downwards in the atmosphere and thus do so
 for  $X_{\max}$ and $X^{\mu}_{\max}$ for a particular shower.
 However, in contrast to $X_{\max}$,  $X^{\mu}_{\max}$  is much less
 sensitive to hadron production   in the primary interaction. The EAS
 muon content rather depends on the multistep hadronic cascade in which the
 number of pions and kaons increases in an avalanche way until the probabilities
 for their decays become comparable to the ones for interactions. 
 For charged pions, this happens when their   energies  approach 
  the  corresponding
 critical energy, $E_{\rm crit}^{\pi ^{\pm}}\simeq 80$ GeV \cite{gaisser}.
  The   maximum  of the muon production profile  is  
  close to this turning point.

As a consequence,  $X^{\mu}_{\max}$ is very sensitive
to the forward spectral shape of secondary mesons in pion-air collisions:
producing in each cascade step a meson of a slightly higher energy would mean
that a larger number of cascade branchings is required for reaching the
critical energy, with the result that the maximum  of the muon production 
profile  will be observed deeper in the atmosphere. 
A similar effect may be produced by a smaller pion-air cross section as this
would increase the pion mean free pass and thereby elongate
the muon production profile.
However, there is another
potential mechanism which may influence model predictions for  $X^{\mu}_{\max}$,
namely, a copious
 production of  baryon-antibaryon pairs in pion-air interactions.
Indeed, (anti-)nucleons do not decay,\footnote{Life time of relativistic neutrons
 exceeds by many orders of magnitude the time scale for EAS development.}  
 hence, they continue to interact even when their energies fall below 100 GeV,
producing additional generations of secondary hadrons in the cascade.
Muons emerging from decays of secondary pions and kaons created in interactions
of such low energy (anti-)nucleons contribute to the elongation of the muon
production profile and give rise to larger values of  $X^{\mu}_{\max}$.
It is noteworthy that the respective effect is noticeable if (and only if)
the yield of  baryon-antibaryon pairs  in pion-air collisions is
 comparable to the one of secondary pions.

For the calculated   $X^{\mu}_{\max}$ (for muon energies
$E_{\mu}\geq 1$ GeV), we observe substantially 
  stronger model dependence than for $X_{\max}$, 
as demonstrated in  Fig.\ \ref{fig:xmax} (right).
  To reveal the physics behind,
 we   use the same ``cocktail'' model
 approach as in  Section \ref{sec:model-xmax}. 
 First, we apply QGSJET-II-04 to describe all the characteristics of the
  primary interaction, while treating the rest of the hadron cascade using 
 either EPOS-LHC or QGSJET, the results shown respectively by the
 blue and green dashed lines in   Fig.\ \ref{fig:xmax} (right). 
  As expected,  the obtained $X^{\mu}_{\max}$ deviates only slightly
 from the original model calculations: the difference between the solid and
 dashed blue lines does not exceed 7  g/cm$^2$, while being even smaller for
 QGSJET (solid and dashed green lines).
Indeed, the bulk of
  the differences between  the model predictions for  $X^{\mu}_{\max}$ 
 is due to secondary  (mostly pion-air) interactions in the cascade. 
  In case of QGSJET, this is mainly caused by somewhat larger inelastic
  pion-  and kaon-air cross sections and softer meson spectra predicted
  by that model.  The first effect is responsible for $\simeq 25$\% of the
  difference between QGSJET and QGSJET-II-04. This is
  illustrated by the green dotted-dashed
  line in   Fig.\ \ref{fig:xmax} (right), 
  which is obtained applying QGSJET-II-04 to 
  describe both the primary interaction and the inelastic cross sections
   for all the secondary hadron-air collisions
in the cascade, while treating hadron production in
secondary hadron-air interactions with QGSJET.
 On the other hand, using  QGSJET-II-04 results also for the
pion and kaon spectra in pion-air collisions   produces an additional 
$\simeq 60$\% effect which thus covers the most of the difference of the
two models' predictions for
 $X^{\mu}_{\max}$, as shown by the green dotted line in    Fig.\ \ref{fig:xmax} (right).

In turn,  the largest part of the difference between EPOS-LHC and  QGSJET-II-04 ($\simeq 35-40$\%)
is due to the copious  production of  baryon-antibaryon pairs in 
the former model. This is illustrated by the blue dotted-dashed line in 
  Fig.\ \ref{fig:xmax} (right), which is obtained applying  QGSJET-II-04 to describe
both the primary interaction and the production of nucleons and antinucleons
in all the secondary pion-air collisions, while treating the rest with
EPOS-LHC.   The  remaining $\simeq 30-35$\%
 difference between the two models is  due to harder
   spectra of secondary mesons in EPOS-LHC for pion-  and kaon-air 
   interactions.
Indeed, using  QGSJET-II-04 results both for the primary interaction and for
hadron spectra in    pion-  and kaon-air collisions,
we obtain the energy dependence  of  $X^{\mu}_{\max}$, shown by the 
blue dotted line in  Fig.\ \ref{fig:xmax} (right), which is very close to 
the pure QGSJET-II-04 calculation.

Finally, let us check the energy range of pion-  and kaon-air collisions
which impact  $X^{\mu}_{\max}$.
As in  Section \ref{sec:model-xmax}, we apply QGSJET-II-04 to treat 
  hadronic  interactions at $E>E_{\rm trans}$,
while describing hadron-air collisions at $E<E_{\rm trans}$ using either
EPOS-LHC or QGSJET. The obtained dependence of  $X^{\mu}_{\max}$ 
on   $E_{\rm trans}$ for the two cases is shown in   
 Fig.\ \ref{fig:xmax-mix} (right)
  by respectively  blue and green dotted lines  for   $E_0=10^{19}$ eV.
Similarly to the  $X_{\max}$ case, the  
range of relevant pion-  and kaon-air collisions extends to very high
 energies,
as reflected by the observed $E_{\rm trans}$-dependence of the calculated
 $X^{\mu}_{\max}$. On the other hand, this energy range is significantly
 broader than in the case of  $X_{\max}$  because all the stages of
 the hadronic cascade development, down to the pion critical energy,
 contribute here.

\section{Conclusions}
\label{sec:conclusions}
We have demonstrated that a substantial part of present uncertainties 
concerning model predictions for the average EAS  maximum 
depth is related to the modeling of  very high energy pion-air collisions.
We traced down the sources of these uncertainties to differences in model
predictions concerning the inelastic  cross sections and
production spectra of mesons and (anti-)nucleons
in pion-nucleus interactions. On the other hand, our
analysis revealed an even stronger sensitivity of the calculated
maximal muon production depth  in air showers to these interaction
characteristics.  Thus, measurements of  $X^{\mu}_{\max}$ 
  by CR experiments have a good potential
 to constrain the treatment of pion-air interactions in the
  very high energy range
 and to reduce thereby model-related uncertainties for $X_{\max}$.
In particular, the results of  the Pierre Auger experiment disfavor
a copious  production of  baryon-antibaryon pairs, predicted by the EPOS-LHC
model, and reduce thereby the range of model uncertainties for  $X_{\max}$. 

In conclusion, $X^{\mu}_{\max}$
 measurements by CR experiments provide an important 
complement to LHC studies for constraining the models of high energy hadronic 
interactions. Further experimental progress in both directions, along with 
improvements in the interaction modeling, will contribute to the resolution of the 
UHECR composition puzzle.

\subsection*{Acknowledgments}
The authors acknowledge useful discussions with T.\ Pierog. 
This work was supported in part by
 Deutsche Forschungsgemeinschaft (project
OS 481/1-1) and  the State of Hesse via the LOEWE-Center HIC for FAIR.

\end{document}